# Reaction Mechanisms of Synthesis of
# 3,4-Epoxybutyric Acid from 3-Hydroxy-γ-Butyrolactone
# by Density Functional Theory

Jong Yu-Chol,   Rim Jong-Won,   Ju Yun-Hui,   Sin Kye-Ryong[*]

( *Faculty of Chemistry, **Kim Il Sung** University, Pyongyang, DPR of Korea*)

*    ryongnam9@yahoo.com

**Abstract:** In this paper, carried out were the investigations on the synthetic reaction mechanisms of 3,4-epoxybutyric acid (EBA) from 3-hydroxy-γ-butyrolactone (HBL) with two different activating agents, methanesulfonyl chloride (MC) or acetic acid (AA), respectively, and on the convertion of EBA to L-carnitine by density functional theory (DFT/B3LYP). The theoretical calculations showed that the two reaction mechanisms of EBA synthesis with MC or AA as an activating agent were nearly the same. If activated HBL is hydrolysed, not only ring cleavage reaction, but also reverse reaction to HBL can take place.   In the case of AA as the activating agent,   the activation energy ( energy barrier ) for EBA synthesis is 1.8 times larger than that with MC. It means that the synthesis of EBA with AA may make more by-products with less yield of EBA than that with MC, that can be one reason why AA gave the less yield than MC in EBA synthesis, as reported in the previous experimental study.

**Introduction**

L-carnitine is known as a powerful natural bioactive substance to protect the bio-functions and promote the energy production in living cells.[1-3] L-carnitine is one of the conditionally essential amino acid that plays an important role as a cofactor in cellular energy production in the mitochondrial matrix. L-carnitine aids in the transport of activated acyl groups across the mitochondrial inner membrane, and it is needed for the oxidation of long-chain fatty acids in the mitochondria of all kinds of cells.[4] L-carnitine exhibits a wide range of biological activities including anti-inflammatory, cardioprotective, gastroprotective, antiapototic and neuroprotective properties.[5-8] Thus it may be used as an aging inhibitor of cells, that is why its synthesis and application is widely researched now.

A dominant step in synthesis of L-carnitine is to synthesize a intermediate substance with epoxy group. Such an intermediate can be synthesized by using epichlorohydrin with epoxy group, but it can cause the difficult problem to separate racemic modifications in the products.

Therefore, the synthetic methods that synthesize EBS from HBL and then convert it directly to L-carnitine without racemic mixtures have become prosperous for L-carnitine synthesis.[9]

To the best of our knowledge, there has not been any theoretical study on the synthetic reaction of 3,4-epoxy butyric acid for L-carnitine, reported up to date.

In this paper, DFT was applied to locate the transition states and provide the mechanisms for EBA synthesis from HBL and discuss the reasons for the different overall yields of L-carnitine, theoretically.

1. Reaction Models and Computational Procedure

Here presented were the reaction models for EBA synthesis from HBL with the different activating agents (MC or AA). ( Fig. 1 ), MGB means 3-methanesulfonyl-γ-butyrolactone, activated by MC and ABG denotes 3-acetate-γ-butyrolactone, activated by AA.

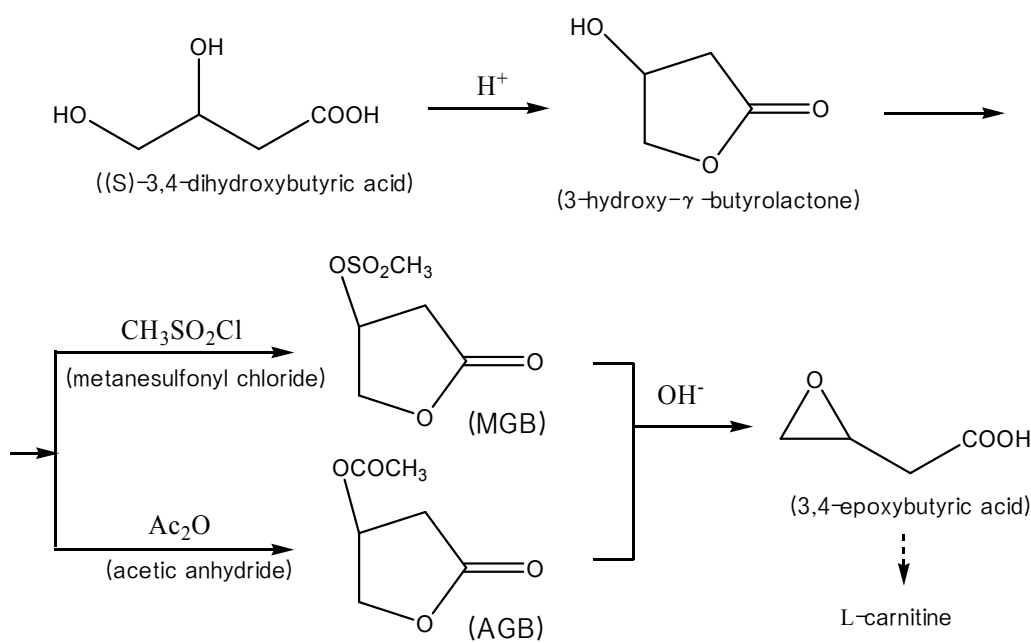

**Fig.** 1   Reaction models for synthetic reaction of L-carnitine

All calculations were carried out by employing SIESTA code [11]. The nonlocal generalized gradient approximation functional (GGA/BLYP[12-14]) in conjunction with the double zeta plus polarization (DZP) orbitals for all electrons in a molecule was used to locate all the possible stationary points for reactant, intermediates, and product in the given reaction pathways. The convergence criteria for these optimizations consisted of threshold values of $2\times10^{-5}$Ha , 0.004 Ha/Å , and 0.005Å for energy, gradient, and displacement convergence, respectively, while a self-consistent field density convergence threshold value of $1\times10^{-5}$Ha was specified. Geometries at the stationary points such as transition state ( TS ) were obtained by using NEB tools in SIESTA.

To take account for several elementary reactions, calculations for the reaction mechanisms were proceeded according to the following steps[15]: first, to confirm the elementary reactions for the synthetic reaction of EBA acid from MGB or AGB. second, to estimate correct pathways of elementary reactions, and then to search optimized TS geometry and calculate the energy barriers on the pathways.

To confirm the elementary reactions, we set MGB or AGB as the initial reactant and EBA as the last product, and composed the reasonable geometry of the intermediates by searching the stationary points on the minimum energy path (MEP) between reactant and product. The trial structures for the intermediate among the possible ones were chosen so as to be distinguished with one another, energetically and geometrically. Then, the elementary reactions were drawn by selecting the pairs of the geometric relatives among these trial intermediates and set them as reactant or product for the given elementary reaction one by one up to arrive at the initial reactant or the last product. For the i-th elementary reaction, the reactant was denoted as $R_M i$ ( M means MGB as the initial reactant ) or $R_A i$ ( A for AGB as the initial ) and the product as $P_M i$ or $P_A i$ as the formers. (Fig. 2)

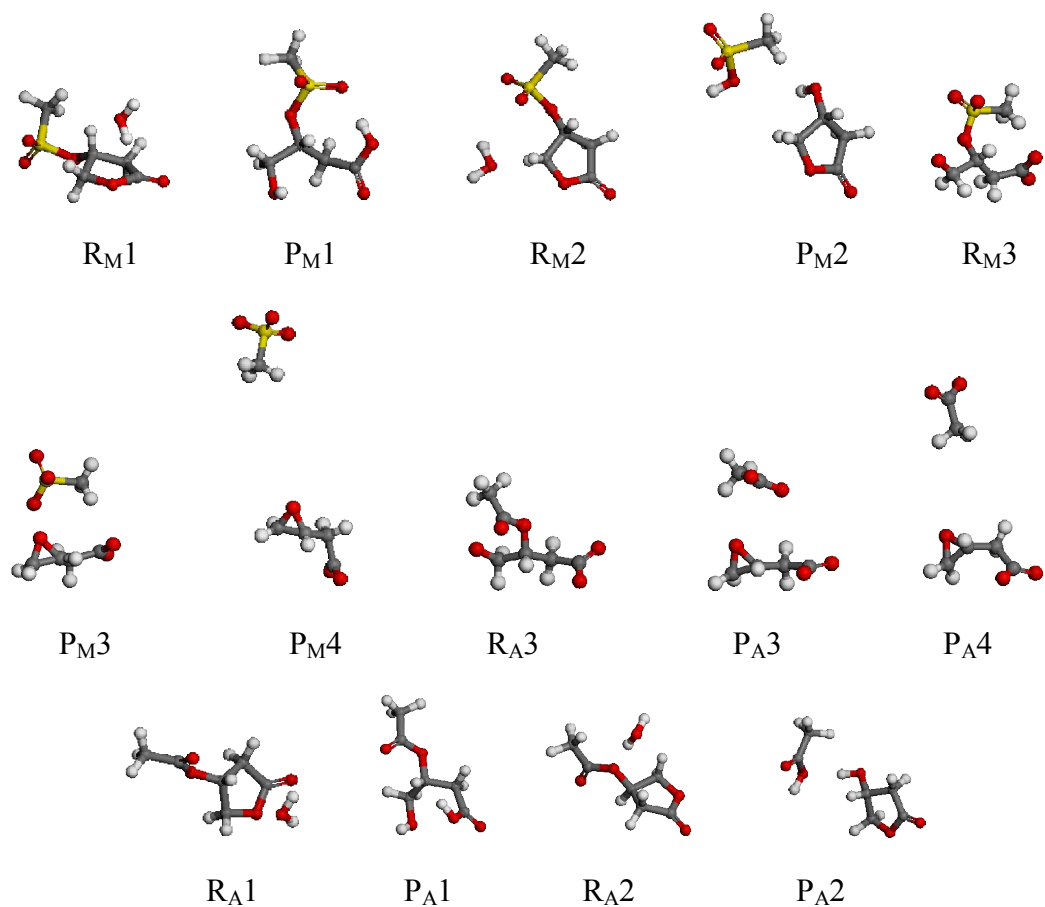

**Fig.** 2  Structural candidates for the intermediates

## 2. Results and Discussion

Some of the elementary reactions composed by using the above chosen intermediates and their TS geometries and energy barriers calculated by DFT were presented in Fig. 3 and Table 1, where $T_Mi$ denotes TS from $R_Mi$.

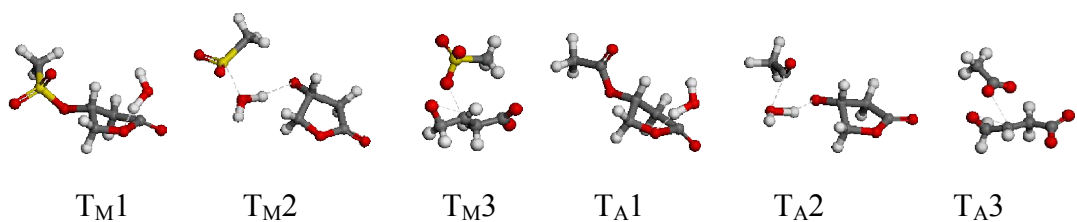

| | $T_M1$ | $T_M2$ | $T_M3$ | $T_A1$ | $T_A2$ | $T_A3$ |

**Fig. 3** The optimized geometries for the transition states

**Table** 1. Elementary reactions and their TS models with activation energy

| No | Elementary reactions | TS models with activation energy |
|---|---|---|
| 1 | $R_M1 \rightarrow P_M1$ (ring cleavage)<br><br>OSO$_2$CH$_3$-lactone + H$_2$O → HO-CH$_2$-CH(OSO$_2$CH$_3$)-CH$_2$-COOH | $T_M1$ (186.4 kJ/mol) |
| 2 | $R_M2 \rightarrow P_M2$ (reverse reaction)<br><br>H$_2$O + OSO$_2$CH$_3$-lactone → OH-lactone + CH$_3$SO$_3$H | $T_M2$ (211.2 kJ/mol) |
| 3 | $P_M1 \rightarrow R_M3$ (dissociation) | |
| 4 | $R_M3 \rightarrow P_M3 \rightarrow P_M4$ (epoxidation)<br><br>$^-$O-CH(OSO$_2$CH$_3$)-CH$_2$-COO$^-$ → epoxide-CH$_2$-COO$^-$ + CH$_3$SO$_2$O$^-$ | $T_M3$ (84.36 kJ/mol) |

| 5 | $R_A1 \rightarrow P_A1$ (ring cleavage) 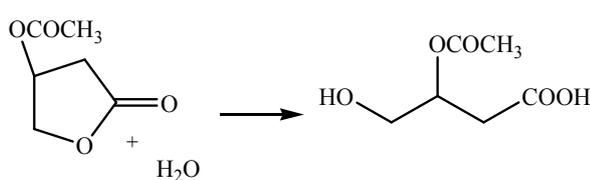 | $T_A1$ (169.4 kJ/mol) 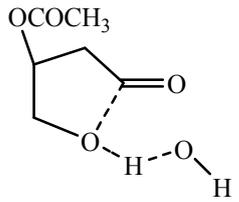 |
|---|---|---|
| 6 | $R_A2 \rightarrow P_M2$ (reverse reaction) 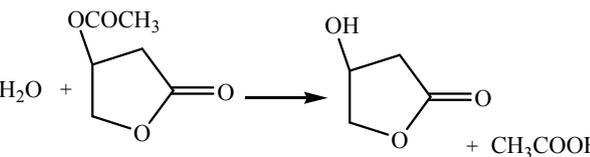 | $T_A2$ (179.5 kJ/mol) 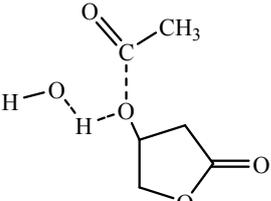 |
| 7 | $P_A1 \rightarrow R_A3$ (dissociation) 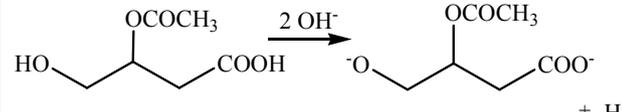 | |
| 8 | $R_A3 \rightarrow P_A3 \rightarrow P_A4$ (epoxidation) 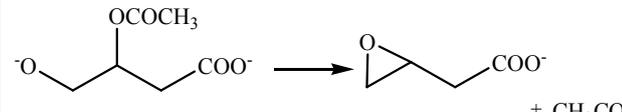 | $T_A3$ (154.5 kJ/mol) 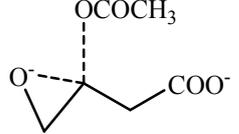 |

In Table 1, the elementary reactions for No.1 and 5 are the ring cleavage reactions where lactone ring was opened by hydrolysis and those for No.2 and 6 are the reverse reactions where the products converted to the reactant, 3-hydroxy-γ-butyrolactone. No.4 and 8 are the epoxidation reactions.

From the elementary reactions in Table 1, the mechanisms of synthetic reaction of EBA from the activated γ-butyrolactones can be set up as follows:

$$P_2 + A \rightarrow HGB$$
$$R_1 + H_2O \rightarrow P_1$$
$$P_1 + 2\,OH^- \rightarrow R_3$$
$$R_3 \rightarrow P_3$$

The schematic energy diagrams of these mechanisms were presented in Fig. 4, where the relative energy is the difference of total energies of the given species from the total energy of $R_M1$ or

$R_A1$ (-2.7466×10$^6$ kJ/mol or -1.6037×10$^6$ kJ/mol, respectively) as zero energy reference.

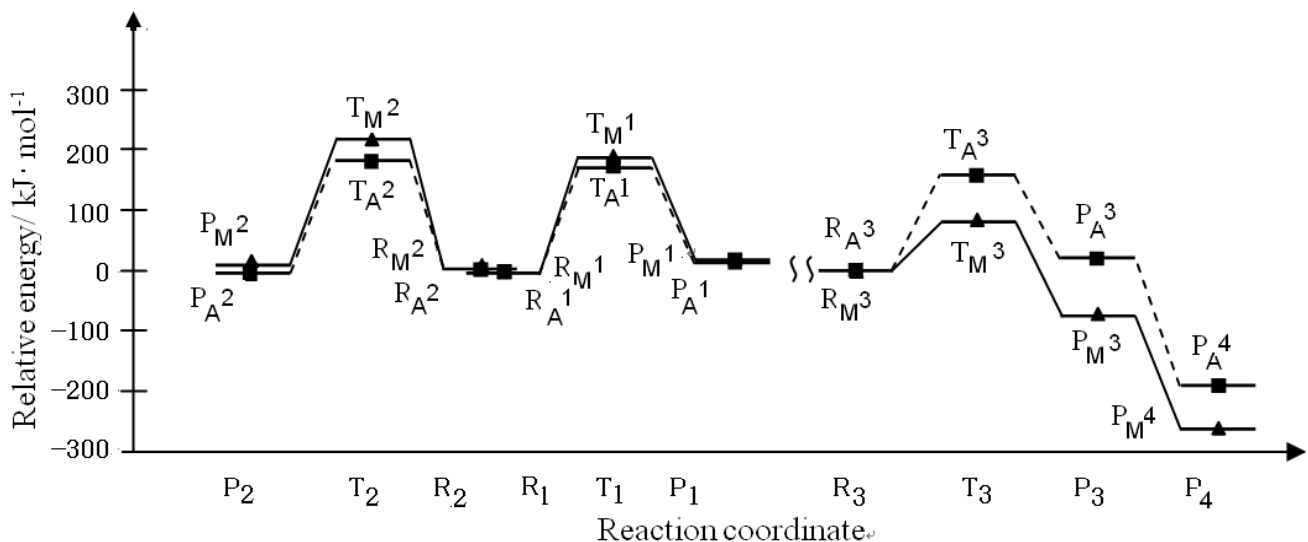

**Fig. 4** The schematic energy diagrams of the mechanisms of synthetic reactions of EBA from the activated γ-butyrolactone by MC or AA

Fig. 4 shows that in case of MC as the activating agent, the activation energy of the reverse reaction (211.2kJ/mol) was larger than that of the ring-cleavage reaction (186.4kJ/mol), but in case of AA, it was 179.5kJ/mol and larger than that of the ring-cleavage reaction (169.4kJ/mol). The synthetic reaction with AA has the activation energy of 154.5kJ/mol, which was 1.8 times larger than that with MC (84.36kJ/mol).

**Conclusions**

By DFT calculations for the reaction mechanisms for synthesis of 3,4-epoxybutyric acid from 3-hydroxy-γ-butyrolactone, it was found out that there was no remarkable difference between the reaction mechanisms with methanesulfonyl chloride or acetic anhydride as activating agent. If the activated γ-butyrolactones were hydrolysed, not only ring cleavage reaction, but also reverse reaction to 3-hydroxy-γ-butyrolactone can take place. In case that the activating agent was methanesulfonyl chloride, the activation energy of the reverse reaction was larger than the ring-cleavage reaction, but in case that the activating agent was acetic anhydride, the activation energy of the reverse reaction was larger than the ring-cleavage reaction. The activation energy with acetic anhydride was 1.8 times larger than that with methanesulfonyl chloride.

It gave the hint that the reaction activated by acetic anhydride may make more by-products and offer less yield than that by methanesulfonyl chloride.